# Mining Time-Stamped Electronic Health Records Using Referenced Sequences

Anne S. Woods is a Senior Statistician, Northern California Institute for Research and Education, San Francisco, CA 94121 (e-mail: anne.woods@va.gov); Craig S. Meyers is a Senior Biostatistician, University of California, San Francisco, Department of Medicine, San Francisco (e-mail: craiger.meyer@gmail.com); Brian C. Sauer is an Associate Professor, Department of Internal Medicine, Division of Epidemiology, Salt Lake City and Principal Investigator, Salt Lake City Veterans Affairs Medical Center, Health Services Research and Development (IDEAS) Center, Salk Lake City, UT 84108 (e-mail: brian.sauer@utah.edu); Beth E. Cohen is an Associate Professor, Department of Medicine, University of California, San Francisco and staff Physician, San Francisco VA Medical Center, San Francisco 94121.

**Abstract:**

Electronic Health Records (EHRs) are typically stored as time-stamped encounter records. Observing temporal relationship between medical records is an integral part of interpreting the information. Hence, statistical analysis of EHRs requires that clinically informed time-interdependent analysis variables (TIAV) be created. Often, formulation and creation of these variables are iterative and requiring custom codes.

We describe a technique of using sequences of time-referenced entities as the building blocks for TIAVs. These sequences represent different aspects of patient's medical history in a contiguous fashion.

To illustrate the principles and applications of the method, we provide examples using Veterans Health Administration's research databases. In the first example, sequences representing medication exposure were used to assess patient selection criteria for a treatment comparative effectiveness study. In the second example, sequences of Charlson Comorbidity conditions and clinical settings of inpatient or outpatient were used to create variables with which data anomalies and trends were revealed. The third example demonstrated the creation of an analysis variable derived from the temporal dependency of medication exposure and comorbidity.

Complex time-interdependent analysis variables can be created from the sequences with simple, reusable codes, hence enable unscripted or automation of TIAV creation.

## 1. Introduction

Electronic Health Records (EHRs) have created large amounts of inexpensive data for health research (Jensen et al. 2010; Coorevits et al. 2013; Murdoch 2013; Yadav et al. 2016; Myers 2016; Casey et al. 2016; Cowie et al. 2017). However, EHRs are designed for various purposes such as administrative record keeping and SQL transaction processes. As a result, systematic biases and confounding relationships in the EHR data must be investigated and accounted for when using EHRs for research (Hripcsak and Knirsch 2011; Rusanov et al. 2014; Casey et al. 2016; Agniel et al. 2018). In addition, patient health history is a continuous picture, while EHRs are discrete records of encounters and transactions. The true state of a patient's health may not be clearly indicated in any single recorded entry; rather it is inferred from combining temporally related records of interventions and outcomes. As it is impossible to predict all sources of bias and confounding in EHRs, it is imperative for the investigators to interrogate the data as easily as possible (Hruby 2016).

Data preprocessing is an integral part of data mining and knowledge discovery (Hand et al. 2001; Kriegel et al. 2007; Maimon and Rokach 2010; Han et al. 2012; Garca et al. Wickham 2014). Kriegel et al. reiterated that this time-consuming step is “more than a tedious necessity”, as “knowledge discovery is more than pure pattern recognition: Data miners do not simply analyze data, they have to bring the data in a format and state that allows for this analysis.”(2006, pp. 91). With respect to mining EHRs, the typical workflow is iterative (Figure 1), involving a human intermediary who is tasked with records extraction and preprocessing under the direction of the researcher. The data preprocessing step is typically labor intensive and performed by the team’s analyst/programmer. Furthermore, this step is performed individually by each team even when the desired analysis variables are the same or similar. Hruby et al. (2016) pointed out that “the current progress in biomedical informatics mainly lies in support for query execution and information modeling”, and  “to realize the promise of EHR data for accelerating clinical research, it is imperative to enable efficient and autonomous EHR data interrogation by end users such as biomedical researchers.”

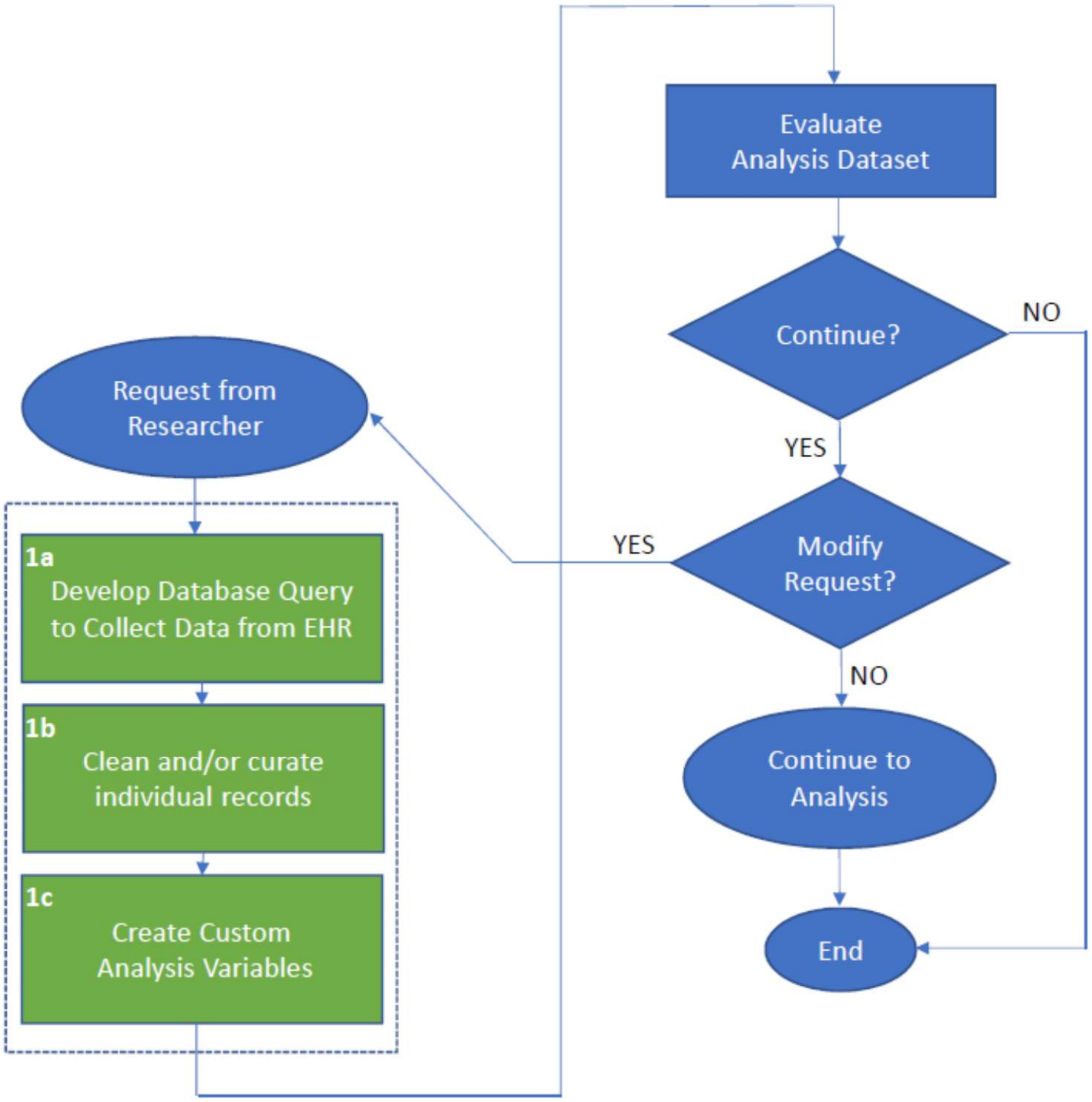

**Figure 1**: Current workflow of a research study using EHRs. Tasks colored in green are typical of the data preprocessing phase. They are usually performed by an analyst or programmer. Tasks colored in blue are performed by the investigator.

Analysis variables are the perspectives and vehicles with which data are analyzed. They are typically the columns of a 2-dimensional input dataset. At the constituent level, all data analyses are calculations performed on suitably chosen analysis variables. Creating analysis variables with EHRs faces a unique challenge in that, while relative timing is an intrinsic part of interpreting medical records, EHRs are time-stamped transactional records. Assembling time-interdependent analysis variables (TIAV) from EHRs requires custom codes that adhere to clinically-informed temporal relationships. At the same time, the development and refinement of the temporal relationship are often informed by the results of previous iterations. The iterative process causes interruption that hinders dynamic and spontaneous exploration of data (Figure 1).

We introduce a technique which enables efficient creation of TIAVs from EHRs. At the core of the technique are sequences of entities representing patient's medical history in a continuous fashion. The entities can be cross-referenced to each other temporally. These sequences are like temporal abstraction (Shahar 1997; Nigrin and Kohane 2000; Post and Harrison 2007; Moskovitch and Shahar 2009; Combi et al. 2012; Lan et al. 2013; Moskovitch et al. 2015; Zhao et al. 2017) in that they are variables whose values incorporate temporal information. They are also different in that temporal abstraction is a TIAV derived from either domain knowledge or an algorithm, while the sequences are intermediate data object between EHRs and TIAVs. In other words, the sequences are building blocks that custom TIAVs can be built from. These sequences act like a reference system connecting snapshots of progression and interactions in patient's medical history.

## 1. Method

We describe this technique in terms of two one-to-one functions: time-relating (f) and state-relating (g). Patient medical history can be measured in different time units such as daily, hourly, or annually. Without loss of generality, we use daily as the natural time unit in our discussion. That is, we would like to obtain the states of a set of medical information of a patient on each day. For instance, we would like to obtain daily values of 4 types of clinical information from a patient cohort: medication 1 exposure (binary), medication 2 exposure(binary), inpatient CPT codes(nominal), and Lab A (numeric) result. For each type of information, there is a function f that maps between calendar date (t) and entry location (k) in the corresponding sequence (Table 1a). For each type of information, there is a function g that maps between the entry value (x) to the clinical information (y) (Table 1b).

| | | **f: 1-to-1 Function between Natural Time and Entry Location in Sequence** | | | |
|---|---|---|---|---|---|
| Entry Location in sequence | Calendar Date | Sequence for Medication 1 Exposure | Sequence for Medication 2 Exposure | Sequence for Inpatient CPT Code | Lab A Result |
| k | t | $f_1$ (k)=t<br>$f_1^{-1}$(t)=k | $f_2$ (k)=t<br>$f_2^{-1}$(t)=k | $f_3$ (k)=t<br>$f_3^{-1}$(t)=k | $f_4$ (k)=t<br>$f_4^{-1}$(t)=k |

**Table 1a:** Time functions for sequences

| | | **g: 1-to-1 Function between Entry Value and Clinical Information** | | | |
|---|---|---|---|---|---|
| Entry Value | Clinical Information | Sequence for Medication 1 Exposure | Sequence for Medication 2 Exposure | Sequence for Inpatient CPT Code | Lab A Result |
| x | y | $g_1$ (x)=y<br>$g_1^{-1}$(y)=x | $g_2$ (x)=y<br>$g_2^{-1}$(y)=x | $g_3$ (x)=y<br>$g_3^{-1}$(y)=x | $g_4$ (x)=y<br>$g_4^{-1}$(y)=x |

**Table 1b:** State functions for sequences

In other words, the entry in any location of a sequence has 2 components of information: when and what. Clinical information at time t can be obtained from $g\left(x_{k=f^{-1}(t)}\right)$, where $x_{k=f^{-1}(t)}$ is the entry location for time t. The time and state functions interpret the sequences. To illustrate, an entry of '10' in a sequence could mean a lab value of 10% on 1/13/2019 or exposure to medication X on 7/15/2018 but not on 7/16/2018.

Through these time and state functions (and their inverses), we can obtain information temporally related to each other in a straight forward manner, as demonstrated in our examples.

The design of the sequence and their implementation are not unique. To illustrate, Figure 2a is a text string of 0's and 1's representing binary (0=No, 1=Yes) indicator for exposure to medication X over 25 days.

| k | 1 | 2 | 3 | 4 | 5 | 6 | 7 | 8 | 9 | 10 | 11 | 12 | 13 | 14 | 15 | 16 | 17 | 18 | 19 | 20 | 21 | 22 | 23 | 24 | 25 |
|---|---|---|---|---|---|---|---|---|---|---|---|---|---|---|---|---|---|---|---|---|---|---|---|---|---|
| x | 0 | 0 | 0 | 0 | 0 | 0 | 0 | 0 | 0 | 1 | 1 | 1 | 1 | 1 | 1 | 1 | 1 | 0 | 0 | 0 | 1 | 1 | 1 | 1 | 1 |

V=text string variable

**Figure 2a:** Sequence as a text string variable representing 25 days of exposure (yes/no) history of medication X

The same information history can be stored as 2 variables (Figure 2b).

| $V_1$ | $V_2$ |
|---|---|
| 10 | 8 |
| 21 | 5 |

V1: start of a 1-segment
V2: length of each 1-segment

**Figure 2b:** Alternative sequence representation

Both constructs contain the same time and state information for each day over the same 25-day period. The computing implementation would depend on the specific construct and the choice depends on the application. They can be simple and transparent as in our examples or opaque to enhance data security.

## 2. Results

We present 3 examples using data from Veteran Administration's Corporate Data Warehouse (CDW) (Fihn et al. 2014) for 1,452,098 veterans who had received a PTSD diagnosis between 2007 and 2015. All data extraction and analyses were performed using SAS Enterprise Guide 7.1 (64 bit). We used text string embodiment of sequences of daily information, and TIAVs were created with common text functions.

### 3.1 Identifying patients and study period based on patterns of medication exposure

In a treatment comparison study, we wish to compare outcomes in patients 1 year before and after 4 medications were added to augment the first-line medication. Specifically, a patient meeting inclusion criteria should have a 2-year period around the date of augmenting medication initiation (index date) such that the patient was 1) free of the index medication ≥ X days during the pre-index year, 2) on the index medication ≥ Y days in a continuous window of Z days during the post-index year, 3) on the first-line medication ≥ W1 and W2 days respectively during the pre- and post- index year, 4) on both augmenting and first-line medications the same day at least U days during the post-index year.

Text strings of '0's and '1's were created for each patient's medications history from 1/1/2007 through 12/31/2015. All strings for the same patient have the same length, with each position representing a single day, where '0' and '1' indicates whether he patient was on the medication that day (1=YES, 0=NO). The first position corresponds to a reference date that varies among patients. Reference date was chosen to be the later of 1) beginning of the study period, and 2) when the patient's pharmacy information first became available. In other words, reference date was the earliest date that information of a filled prescription was expected to be in EHR. For example, if the patient's medication record started before the study period, their reference date would be 1/1/2007. However, if the patient's medication records started on 6/15/2014, his reference date would be 6/15/2014. All strings end on 12/31/2015.

Prescription information was translated into series of ‘0’s and ‘1’s by the prescription release dates and days’ supply and allowed storing of unused medications when a prescription was refilled before the previous supply had run out. For instance, two 30-day prescriptions released on 7/30/2010 and 8/25/2010 were represented by 60 ‘1’s from 7/30/2010 through 9/27/2010. Figure 3 shows the time and state functions of the medication sequence.

- $f(k) = \begin{cases} reference\ date, & if\ k = 1 \\ reference\ date + (k-1), & if\ k > 1 \end{cases}$
- $f^{-1}(t) = \{t - (reference\ date) + 1$

- $g(x) = \begin{cases} on\ medication, & if\ x = 1 \\ not\ on\ medication, & if\ x = 0 \end{cases}$
- $g^{-1}(y) = \begin{cases} 1 & if\ y = on\ medication \\ 0, & if\ y = not\ on\ medication \end{cases}$

**Figure 3:** Time and state functions of the medication sequence

To assess degrees of same-day medication overlap (criteria 4), composite binary strings were created from the 2 individual medication strings by comparing entries of the same position. For the composite string, ‘1’ and ‘0’ indicate whether both medications were taken on the same day. Text strings of a hypothetical patient are shown in Figure 4.

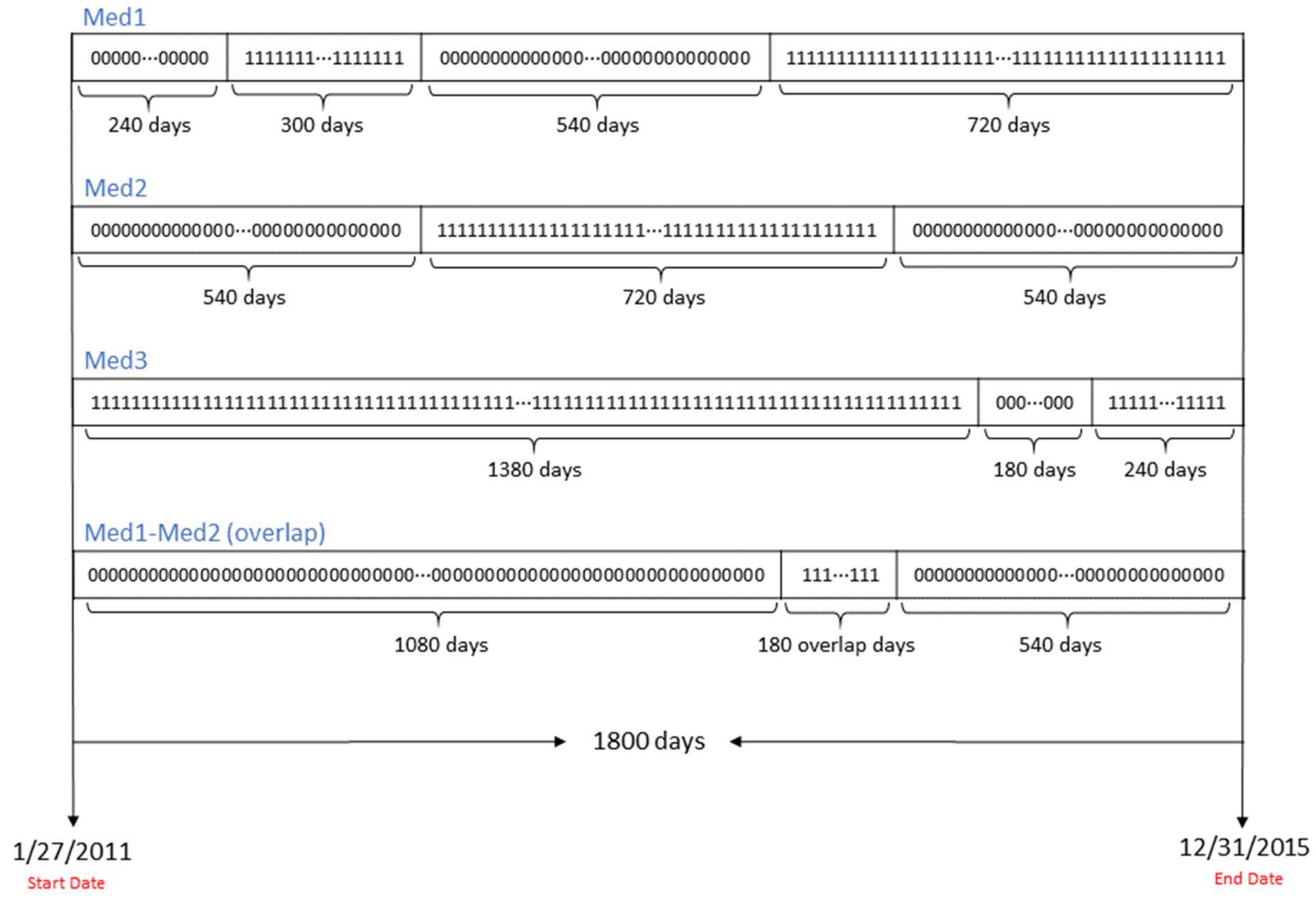

**Figure 4:** Text strings representing use patterns of 3 medications and the overlap between medications 1 and 2 for a hypothetical patient

Several TIAVs were created to evaluate different formulations of the selection criteria (Table 2).

| Time-Depending Analysis Variable | Variable Type | Functions Used |
|---|---|---|
| X: Days of continuous wash-out period | Numeric | PRXMATCH |
| Y: Days on index medication within window of Z days | Numeric | SUBSTR, COMPRESS, LENGTHN |
| W1: Days on first-line medication during pre-index year | Numeric | SUBSTR, COMPRESS, LENGTHN |
| W2: Days on first-line medication during post-index year | Numeric | SUBSTR, COMPRESS, LENGTHN |
| U: Days on first-line and index medications during post-index year | Numeric | SUBSTR, COMPRESS, LENGTHN |

**Table 2:** Analysis variables created to evaluate patient eligibility

Different cutoff values for X, Y, Z, W1, W2 were applied against the respective TIAVs to determine patient eligibility and sample size.

The final selection criteria were that patient 1) should have had at least 180 days free of the index augmenting medication prior to the start of the regimen, 2) was prescribed the index medication ≥ 60 days in a 120-day period during the post-index year, 3) was prescribed the first-line medication ≥ 30 and ≥ 60 days during the pre- and post-index year, respectively. Minimum days on both first-line and augmenting medication were not required. Figure 5 depicts an example of strings that satisfy these selection criteria.

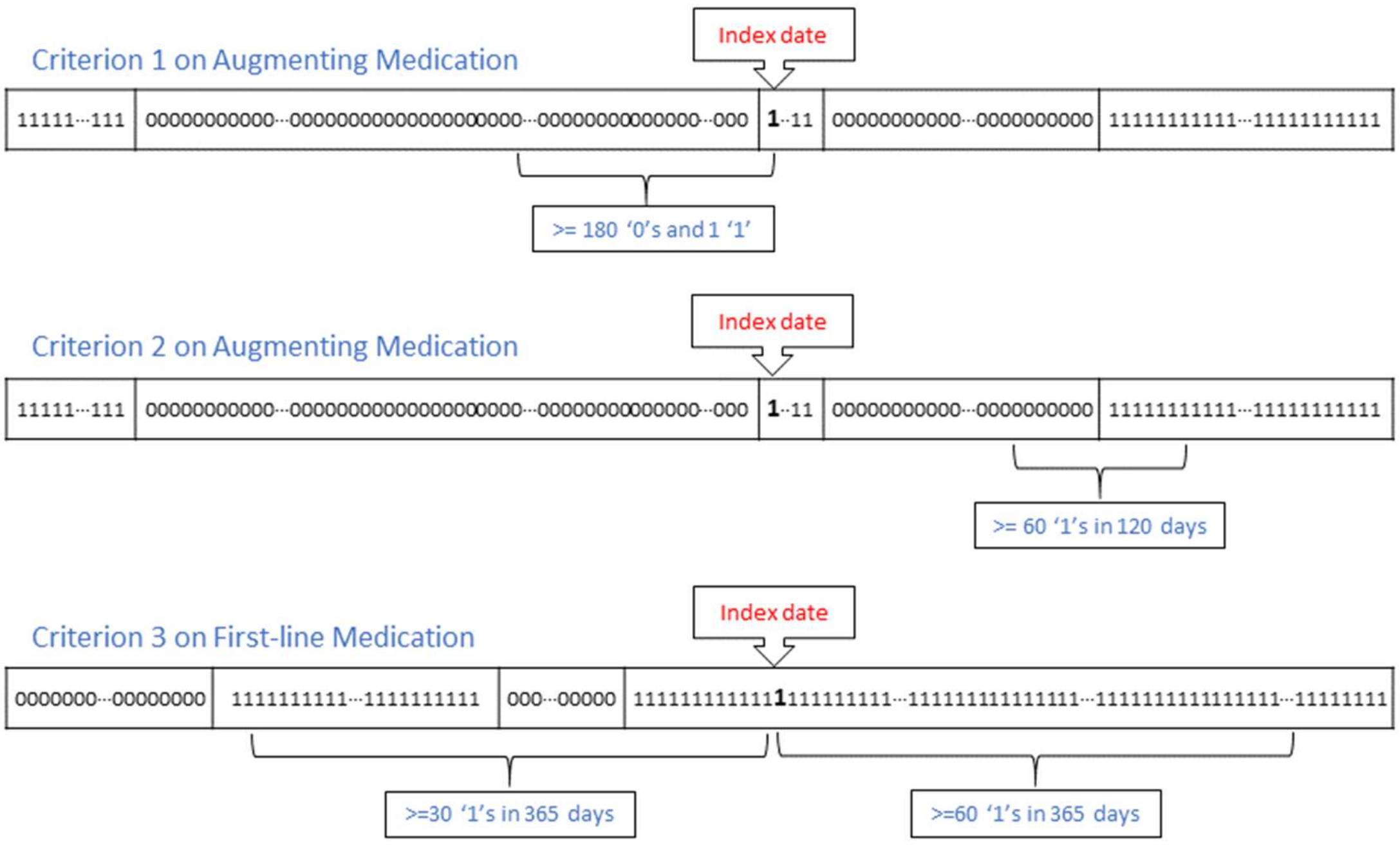

**Figure 5:** Final selection criteria based on use patterns of augmenting and first-line medications

### 3.2 Charlson Comorbidities and Clinic Settings of Diagnosis

Patient comorbidities are clinical elements included in many medical research studies. In this example, 2 strings were created for each patient: comorbidities and diagnosis settings. Numbers 1 to 9 and letters A to H were used to represent the 17 comorbidities in Quan's Enhanced Charlson Comorbidity Index (Quan et al. 2005) based on ICD9 codes (Table 3). Letters 'I' and 'O' represent, respectively, inpatient and outpatient setting in which a diagnosis was given. This example used data prior to the VA transition to ICD 10 codes, but methods could be similarly adapted for ICD 10. The reference point is the start date, which was the earlier of the earliest diagnosis date and 1/1/2007.

| Symbol | Comorbidity |
|---|---|
| 1 | Myocardial infarction |
| 2 | Congestive heart failure |
| 3 | Peripheral vascular disease |
| 4 | Cerebrovascular disease |
| 5 | Dementia |
| 6 | Chronic pulmonary disease |
| 7 | Rheumatic disease |
| 8 | Peptic ulcer disease |
| 9 | Mild liver disease |
| A | Diabetes without chronic complication |
| B | Diabetes with chronic complication |
| C | Hemiplegia or paraplegia |
| D | Renal disease |
| E | Any malignancy, including lymphoma and leukemia, except malignant neoplasm of skin |
| F | Moderate or severe liver disease |
| G | Metastatic solid tumor |
| H | AIDS/HIV |

**Table 3:** 17 Conditions from the Charlson Comorbidity Index

Of the 1.4 million patients, 860,826 of them had at least one of the 17 conditions. Patients were grouped into separate cohorts by the year of their earliest health records.

For the same patient, strings of comorbidity and clinical setting had the same length, representing start date to 9/30/2015. If a patient received different comorbidity diagnoses on the same date, they are represented in the same position but on a separate string, i.e. string block (Figure 6).

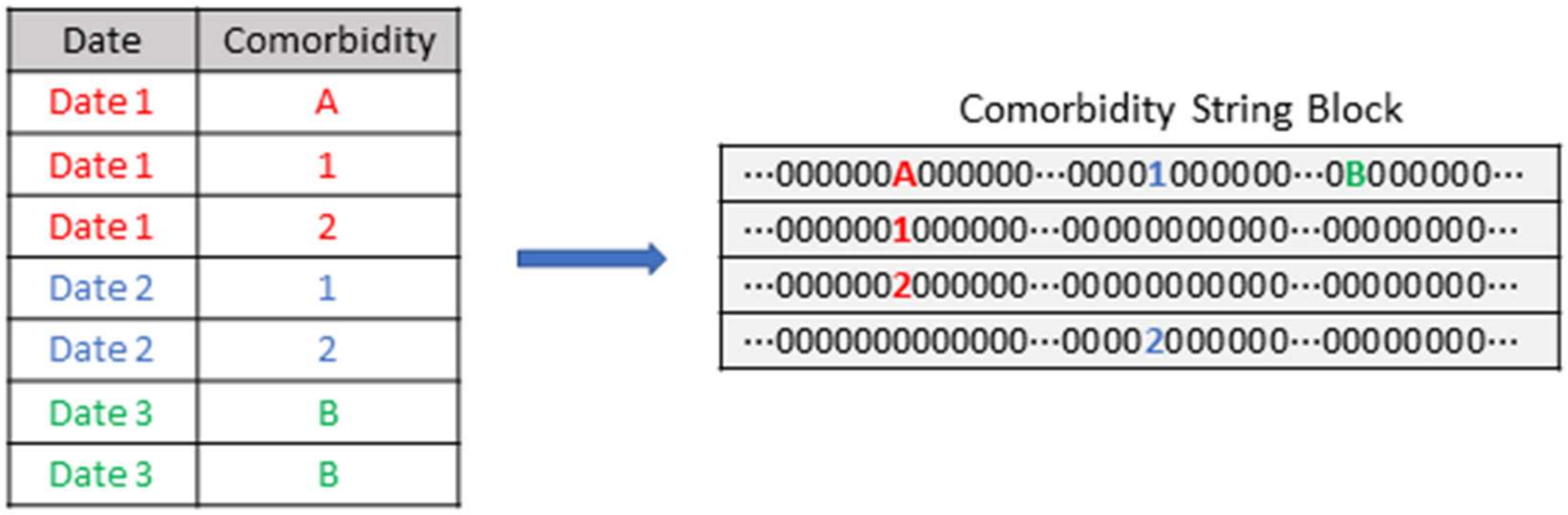

**Figure 6:** String block representing comorbidities diagnosed on 3 dates. Different comorbidities diagnosed on the same date are on separate strings

The Charlson Comorbidity Index (CCI) was simply the weighted sum of the 17 binary variables (Supplemental). Figure 7 depicts the average CCI by length of measurement period, from 365 to 3195 days, at quarterly increments.

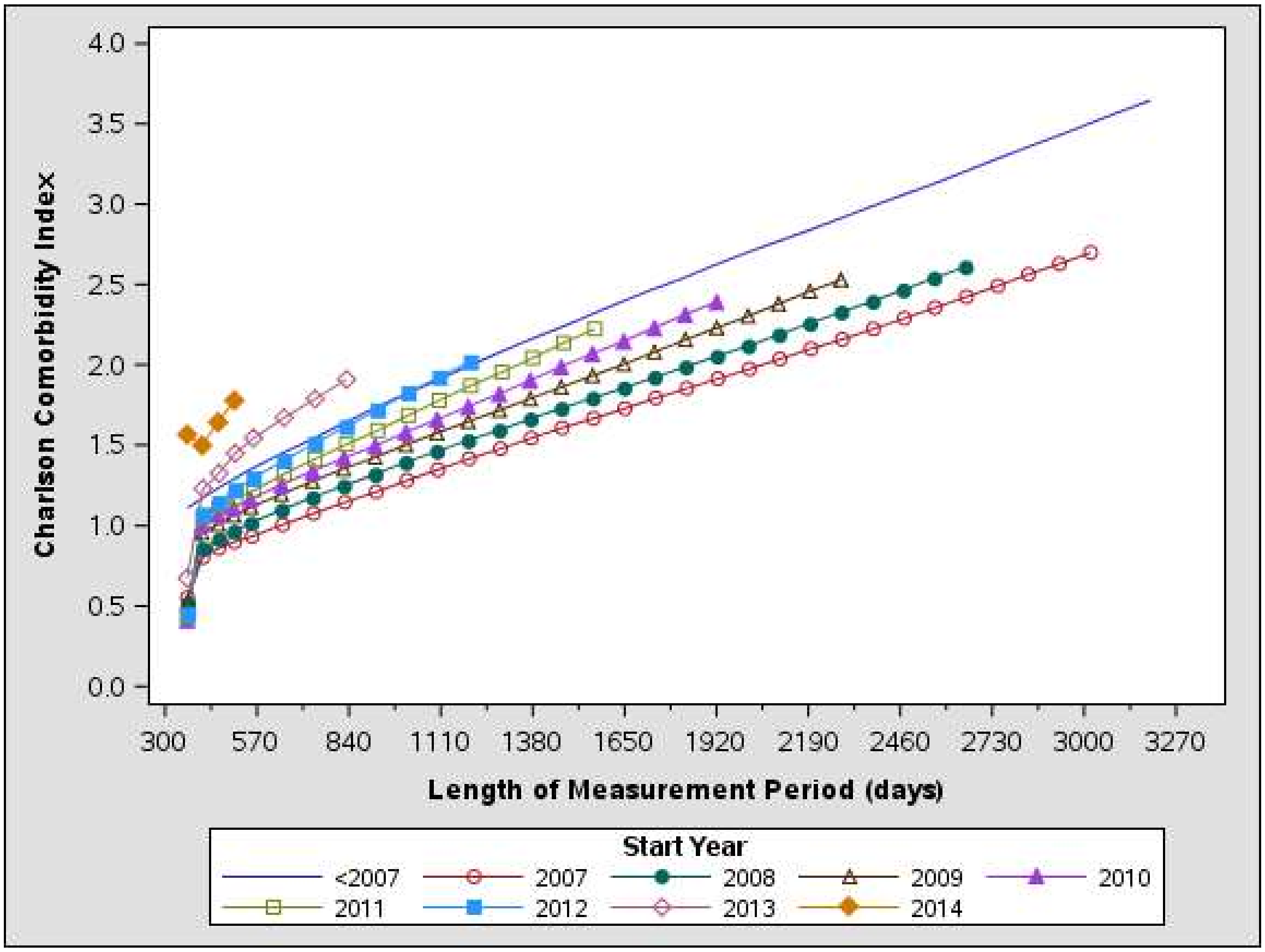

**Figure 7:** Charlson Comorbidity Index increases with length of measurement period

As seen in Figure 7 there is an overall positive association with the length of measurement period. This is not surprising as one would expect both aging and longer observation time frame contributed to the increasing CCI. However, the low initial CCI in cohorts 2007-2012 warranted further investigation. The same calculation was repeated while fixing the measurement period to 365 days immediately before each quarterly report date. We hypothesized that the initial low CCI might be due to the artificially low utilization, as new patients gaining familiarity with VA healthcare system may not have had as many provider visits as established patients. A TIAV that represent total count of ‘I’ and ‘O’ in clinical setting substrings corresponding to each measurement period were created. These counts were proxy measurements for the extent of health service utilization. Figure 8 shows the quarterly plot of CCI by each cohort of start year. Data labels are the median days of service utilization. For cohorts 2007-2012, the initial utilization was dramatically lower.

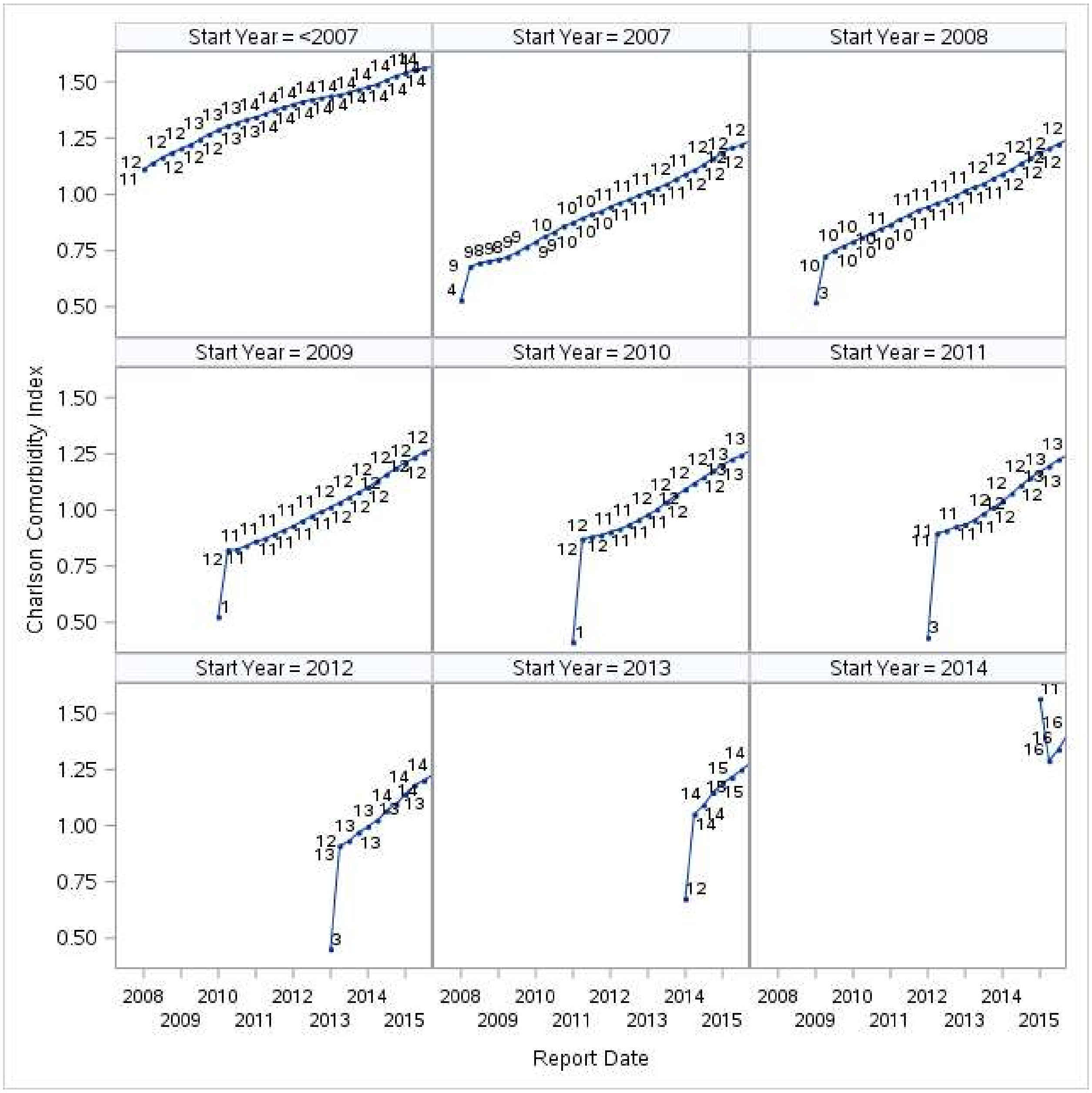

**Figure 8:** Charlson Comorbidity Index from 1-year measurement period. Data labels are measurements of health service utilization (median number of diagnosis dates)

### 3.3 Single Patient Medication Histories and Diagnoses

It's often helpful to view clinical interventions and outcomes together. Figure 9 depicts continuous picture of a hypothetical patient's exposure to 4 classes of medications in a 5-year period, during which the patient experienced 2 myocardial infarctions (MI).

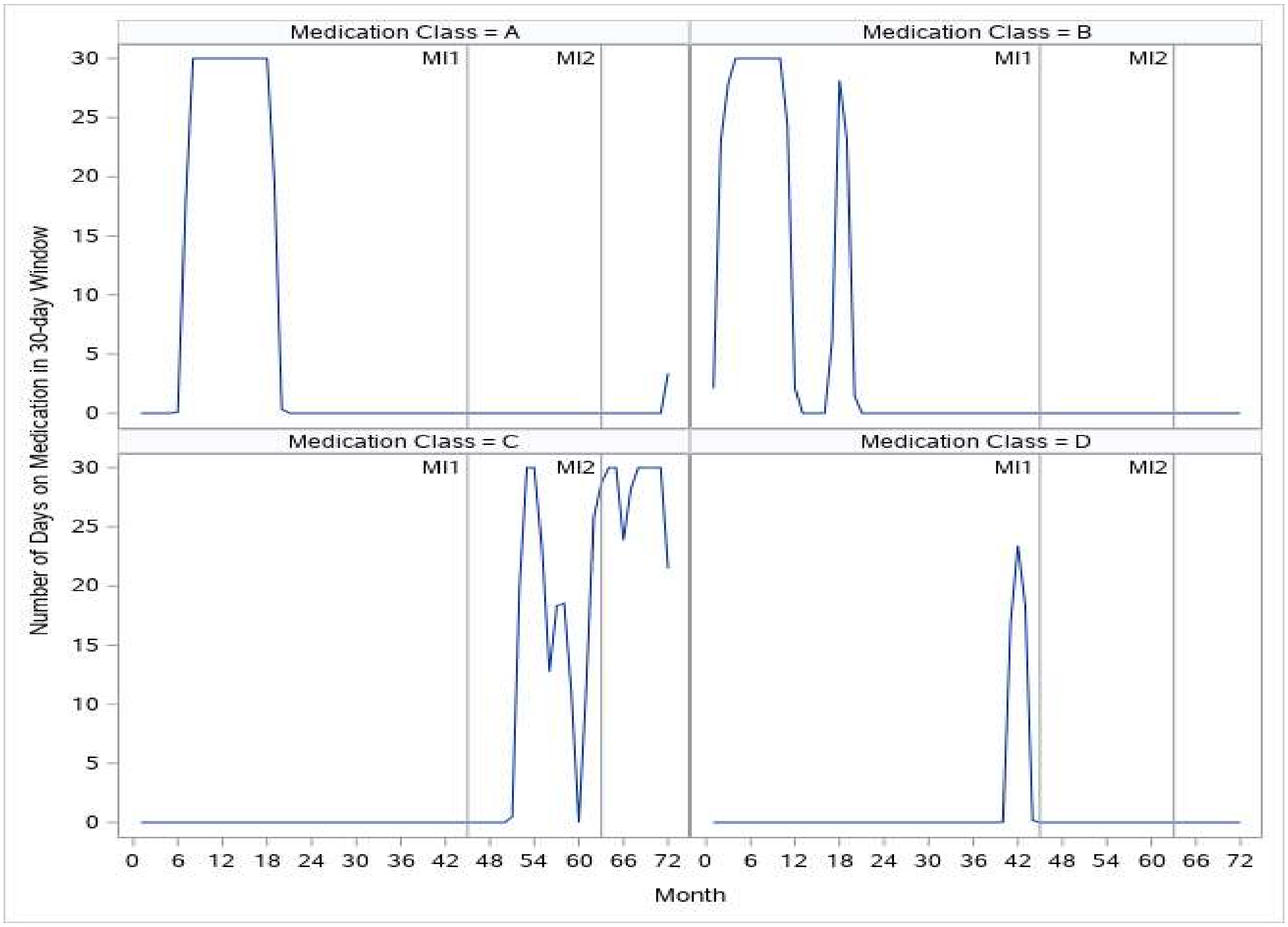

**Figure 9:** 30-day moving averages of medication exposures and MI diagnosis of a hypothetical patient

More generally, when a patient is on multiple medical regimens, it is challenging to tease out the effect of any specific treatment. One approach is to use time-varying covariates to account for the variation associated with the timing and exposure of the treatments. Creation of these variables from time-stamped records require complex custom codes. In contrast, they can be created with the sequence technique in a straight forward manner.

The first step of creating a time-varying covariate is to translate outcome measurement dates into position numbers in the string. These positions 'stake' the search segments. As an illustration, Figure 10 depicts the 30-day windows prior to each of the 4 measurement dates.

Time-varying covariates can be used to investigate possible confounding relationships in EHRs. Incorporating suitable time-varying covariates can also improve the accuracy and precision of the statistical model.

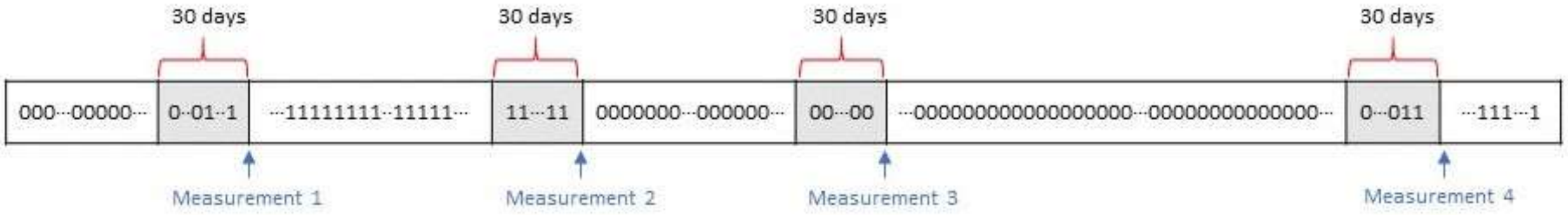

**Figure 10:** 30-day windows prior to each measurement dates

## 3. Discussion

Analysis variables are the vehicles with which to investigate data. Intrinsic to mining EHRs is to observe their temporal relationship among interventions and outcomes.  However, it is time consuming to weave together many time-stamped records to create variables which capture complex temporal dependencies. This challenge hinders spontaneous and thorough investigation of EHRs.

We demonstrate a technique of creating complex analysis variables using a system of referenced entities derived from EHRs. These sequences are not alternative forms for storing EHRs, nor do they function like standards. Instead, they are intermediate data objects that facilitate faster creation of time interdependent analysis variables. These sequences can be utilized by many and do not take up excessive storage space. In our medication strings example, string data took up 0.44 GB of storage as compared to 3.4 GB of prescription-level data. For the comorbidity example, 3.2 GB of source records were condensed into 0.28 GB of string data. More importantly, in share resource environment like Veterans Informatics and Computing Infrastructure (VINCI), there is intense pressure on programming efficiency to reduce the impact on others working in the collective environment. Using these sequences as the common data model can greatly reduce the amount of network traffic, computation resources, and programming time.

Kriegel and colleagues envisioned the future of data preprocessing to be “faster and more transparent”, and that “A common data representation and a common description language for data preprocessing will make it easier for both computer and data miner” (2006, pp. 92). We put forth some ideas that could further broaden the method’s utility.

Large EHR databases like the VA CDW support hundreds of research projects. In our own experience, this technique has enabled us to answer ad hoc questions and evaluate new hypotheses with increased efficiency. A library of sequences, created for all VA patients over their entire medical histories, would enable more researchers to scale down data preparation tasks and focus more on analysis. This is because the programming involved in manipulating sequences into TIAVs is simpler, more generalizable, hence more amenable to automation. Application interfaces could be developed to create and summarize a variety of TIAVs. This would enhance the autonomy domain experts to contemporaneously query EHRs in a feedback loop.

The process of standardization reduces subjectivity in the initial cleaning before transforming the data into referenced sequences. Sequences created from standardized data or common data models such as Observational Medical Outcomes Partnership (OMOP)(Overhage 2012) could further expand user group and comparability of results.

Interactive visualization that displays simultaneously a patient's treatment and outcome histories could be a useful clinical tool (Plaisant et al. 1998; Hirsch et al. 2015). Since the sequences are already arranged in a sequential fashion, they are a good data structure for such interface. Additionally, point-and-click capability that brings up additional details in source records of selected entities could be useful for individual patient care.

## 4. Conclusion

We demonstrate a method that can facilitate the transformation of time-stamped EHRs into time-interdependent analysis variables without repeatedly processing the transaction-level data. At the core of the method are sequences containing different types of information from a patient's health history. These sequences can be inter-related to each other temporally through time and state functions. Together, they form a network of information which enable direct access to information in any time span. This data structure is more suitable for streamlined or standardized programming than time-stamped records.

By enabling more efficient and spontaneous creation of custom analysis variables, more efforts can be focused on investigating patterns and insight from the ever-expanding EHRs

**Acknowledgement:**

This work was partially supported by Patient Centered Outcomes Research Institute Grant 2015C2-150-31834 and Department of Defense Grant PR151250. The authors thank Benjamin Glicksberg and Kevin Martin for their helpful comments.